\documentclass[12pt]{article}
\usepackage{epsfig}  
\def\bea{\begin{eqnarray}} \def\eea{\end{eqnarray}} \def\P{{\cal P}} \def\d{{\rm d}}
\def\w{\u{u} }  
\begin{document}
\title{\bf Angula distanco en kaloto \\}  
\author{ A.F.F. Teixeira \thanks{teixeira@cbpf.br} \\
         {\small Centro Brasileiro de Pesquisas F\'{\i}sicas } \\ 
         {\small 22290-180 Rio de Janeiro-RJ, Brazil} } 
\date{2$^a$ de Decembro de 2003}
\maketitle 
\abstract{%c:/Teixeira/ArqTex/Kaloto.tex 
A cap in a 2-D sphere is considered, smaller than an hemisphere. Two points are randomly chosen in the cap. The probability density that the angular separation between the points have a given value is obtained. The knowledge of this function is important for some methods for determining the shape of the universe. 
\vskip1cm 
Kaloto de 2-D sfera estas studata, pli malgranda ol duonsfero. Du punktoj estas hazarde elektataj en la kaloto. La probabla denso de angula distanco inter la punktoj estas havita. Scio de tiu funkcio estas grava por kelkaj metodoj por specifi la formo de l'universo.   
}
\section{Anta\u uparolo} %%%%%%%%%%%%%%%%%%%%%%%%%%%%%%%%%%%%%%%%%%%%%%%
Tre antikva demando estas: kiu estas la formo de l'universo? 
Por havi respondon, precizaj observoj kaj mezuroj de kosmaj objektoj estas necesaj. 
Hodia\u ue, multaj plibonigoj estas okazantaj en la rilata teknologio. 
El pluraj aliaj, la metodo PASHpc studas angula distanco inter kosmaj objektoj de \^ciela arka\^{\j}o \cite{BernuiThyrso}.  
Por esti vere efika, tiu metodo bezonas uzi ekzaktan formon de funkcio EPASHpc. 
En nia artikolo, ni findas tiun funkcion, kiun ni skribas $\P(\alpha,\beta)$. 
  
\section{Esprimo de problemo} %%%%%%%%%%%%%%%%%%%%%%%%%%%%%%%%%%%%%%%%%%
Pensu pri sfero $S^2$ kun radio=1. 
Konsideru kaloton kun angula duondistanco $\alpha\leq\pi/2$ radianoj (se $\alpha=\pi$, la kaloto estus la tuta sfero). 
Elektu hazarde du punktojn de la kaloto.  
Ni demandas, kiu estas la probablo $\P(\alpha,\beta)$d$\beta$ de la angula distanco inter la du punktoj havu valoron inter $\beta$ kaj $\beta+$d$\beta$. 
Certe $0\leq\beta\leq2\alpha$, kaj la normala kondi\^co devigas  
\bea                                                                \label{norm} %%1
\int_0^{2\alpha}\P(\alpha,\beta)\d\beta = 1. 
\eea 

Ni montros ke  
\bea \nonumber
\P(\alpha,\beta)=2\kappa\sin\beta\Big[\sin^{-1}\left(\frac{(\cos\alpha+\cos\beta)\sqrt{\cos\beta-\cos(2\alpha)}} {\sin\alpha(1+\cos\alpha)\sqrt{1+\cos\beta}}\right) 
\eea \vskip-12mm 
\bea \eea \vskip-12mm                                              \label{Pbmea1}  %%2
\bea \nonumber 
+(1-2\cos\alpha)\cos^{-1}\left(\frac{\cos\alpha\sin(\beta/2)}{\sin\alpha\cos(\beta/2)}\right)\Big].
\eea 
Numeraj kalkuloj forte sugestas valoron $1/\kappa=2\pi(1-\cos\alpha)^2$, sed ni ankora\u u ne sukcesis analitike montri tion. 

\section{Solvo de problemo}  %%%%%%%%%%%%%%%%%%%%%%%%%%%%%%%%%%%%%%%%%%%%%%%%%%% 
Ni hazarde elektas punkto $P$ en la kaloto.  
La probablo ke $P$ estas je angula distanco inter $\gamma$ kaj $\gamma + $d$\gamma$ de la centro $O$ de la kaloto estas proporcia al sin$\gamma$. Certe $0\leq\gamma\leq\alpha$. 

Nun ni hazarde elektas novan punkton $Q$ en la kaloto.  
La probablo ke $Q$ estas je angula distanco $\beta$ de $P$ estas proporcia al la arko de cirklo  (kun radio sin$\beta$) kun centro $P$, kaj tute entenata en la kaloto.  
Vidu bildon 1. 

\vspace*{3mm}
\centerline{\epsfig{file=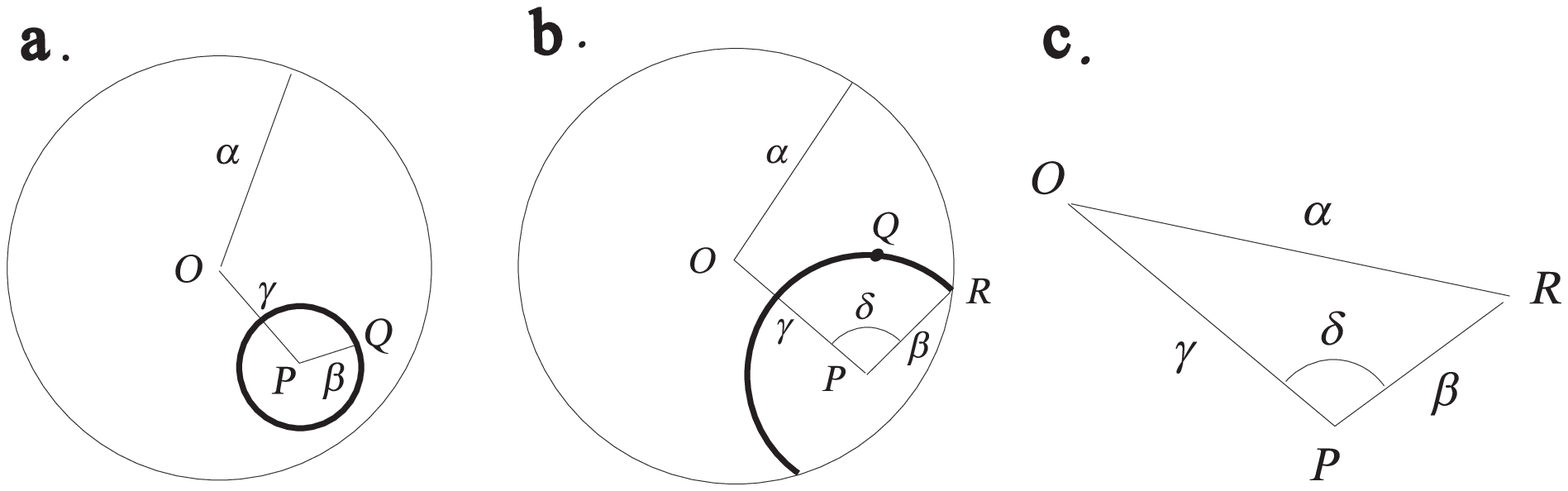,width=12cm,height=36mm}} 
\vspace*{3mm} 

\noindent {\small {\bf Bildo 1} {\bf a}. {\it Kiam $\alpha>\beta+\gamma$, la {\rm locus} de punkto $Q$ estas cirklo} $2\pi{\rm sin}\beta$ {\it longa}. \\ 
{\bf b}. {\it Kiam $\alpha<\beta+\gamma$, la {\rm locus} de $Q$ estas arko de cirklo} $2\delta\sin\beta$ {\it longa}. \\
{\bf c}. {\it En la sfera triangulo $OPR$ validas ekvacio} (\ref{del}). 
}  
\vspace*{5mm}

Se $2\delta$ estas la angulo de la arko, ni havus  
\bea                                                                  \label{del} %%3
\cos\alpha=\cos\beta\, \cos\gamma\, +\, \sin\beta\, \sin\gamma\, \cos\delta. 
\eea

La kunmeta probablo, de $P$ esti en $\gamma$ {\bf kaj} $Q$ esti en $\beta$, estas  
\bea                                                                  \label{Pabc} %%4
\P(\alpha,\beta,\gamma)\d\beta\,\d\gamma = \kappa\Big[2\pi\Theta(\alpha-\beta-\gamma) + 2\delta\Theta(\beta+\gamma-\alpha)\Big]\,\sin\beta\,\sin\gamma\,\d\beta\,\d\gamma;  
\eea 
tie $\Theta$ estas la \^stupo funkcio kun valoroj 0 kaj 1. 
Devas okazi  
\bea                                                                   \label{norm2} %%5
\int_0^\alpha\P(\alpha,\beta,\gamma)\d\gamma = \P(\alpha,\beta), 
\eea 
kaj la konstanto $\kappa(\alpha)$ devos fari veran la (\ref{norm}). 

Ni unue studas la cirkonstanco $\beta<\alpha$. 
Ni kalkulu  
\bea \nonumber 
\P(\beta<\alpha)=2\kappa\pi\sin\beta\int_0^{\alpha-\beta}\sin\gamma\d\gamma 
\eea 
\vskip-11mm \bea \label{P} \eea \vskip-11mm                                          %%6
\bea \nonumber 
+ 2\kappa\sin\beta\int_{\alpha-\beta}^\alpha\sin\gamma\cos^{-1}\left(\frac{\cos\alpha-\cos\beta\,\cos\gamma}{\sin\beta\,\sin\gamma}\right)\d\gamma.
\eea
La unua integralo estas simpla,  
\bea                                                                    \label{P1} %%7
\P_1(\alpha<\beta)=2\kappa\pi\sin\beta\Big[1-\cos(\alpha-\beta)\Big]. 
\eea 
Por kalkuli la duan integralon ni \^san\^gas la variablo $\gamma\rightarrow x=\cos\gamma$ kaj skribas  
\bea                                                                \label{P2}     %%8
\P_2(\alpha<\beta)=2\kappa\sin\beta\int_{\cos\alpha}^{\cos(\alpha-\beta)}\cos^{-1}\left(\frac{\cos\alpha-x\cos\beta}{\sin\beta\sqrt{1-x^2}}\right)\d x.
\eea
Poparta integralado donas   
\bea \nonumber                                                                      
\P_2(\alpha<\beta)=2\kappa\sin\beta\Big[x\cos^{-1}\left(\frac{\cos\alpha-x\cos\beta}{\sin\beta\sqrt{1-x^2}}\right)\,   
- \cos\alpha\sin^{-1}\left(\frac{x-\cos\alpha\cos\beta}{\sin\alpha \sin\beta}\right) 
\eea \vskip-5mm
\bea                                                                  \label{Pbmea} %%9                                                                
+ \frac{1}{2}\tan^{-1}\left(\frac{(1+\cos\alpha\,\cos\beta)(1+x)-(\cos\beta+\cos\alpha)^2}{(\cos\beta+\cos\alpha)\,R}\right) 
\eea \vskip-5mm
\bea \nonumber
 +\frac{1}{2}\tan^{-1}\left(\frac{(1-\cos\alpha\,\cos\beta)(1+x)-
(\cos\beta-\cos\alpha)^2}{(\cos\beta-\cos\alpha)\,R}\right) 
\Big]_{x=\cos\alpha}^{x=\cos(\alpha-\beta)} ; 
\eea 
tie  
\bea                                                                      \label{R}  %%10
R:=\sqrt{[x-\cos(\alpha+\beta)][\cos(\alpha-\beta)-x]}.
\eea
Ni substituas limitojn de integralado kaj adicias la $\P_1(\alpha<\beta)$ de (\ref{P1}), kaj fine findas la (\ref{Pbmea1}).

Nun ni studu cirkonstancon $\beta>\alpha$. 
La denso de probablo venos el  
\bea                              					    \label{Pbmaa}  %%11
\P(\beta>\alpha)=2\kappa\sin\beta\int_{\beta-\alpha}^\alpha\sin\gamma\cos^{-1}\left(\frac{\cos\alpha-\cos\beta\,\cos\gamma}{\sin\beta\,\sin\gamma}\right)\d\gamma.
\eea 
Rimarku ke (\ref{Pbmaa}) estas simila al la dua integralo de (\ref{P}). 
Do ni denove \^san\^gas variablon $\gamma\rightarrow x=\cos\gamma$, kaj denove poparto integralado  donas esprimojn (\ref{Pbmea}) kaj (\ref{R}). 
Ni simpligas \^ciujn esprimojn, nun kun $\beta>\alpha$, kaj findas ke (\ref{Pbmea1}) estas ankora\u u valida. 

Kiam $\alpha$ estas tre malgranda okazas, el (\ref{Pbmea1}), 
\bea                                                               \label{plano}  %%12
\P(\alpha\rightarrow0,\beta)=\frac{8}{\pi\alpha}\frac{\beta}{2\alpha}\Big[\cos^{-1}\frac{\beta}{2\alpha}-\frac{\beta}{2\alpha}\sqrt{1-(\beta/(2\alpha))^2}\Big]; 
\eea
tie ni konsideris $\kappa=2/(\pi\alpha^4)$. \^Ci tiu estas la bone konata rizulto por plana disko \cite{membranas}.
Kaj kiam $\alpha=\pi/2$, kaloto estante la tuta duonsfero, ni havas  
\bea                                                           \label{hemisferio} %%13
\P(\pi/2,\beta)=(1-\beta/\pi)\sin\beta. 
\eea

Vidu bildon 2. 

\vspace*{3mm}
\centerline{\epsfig{file=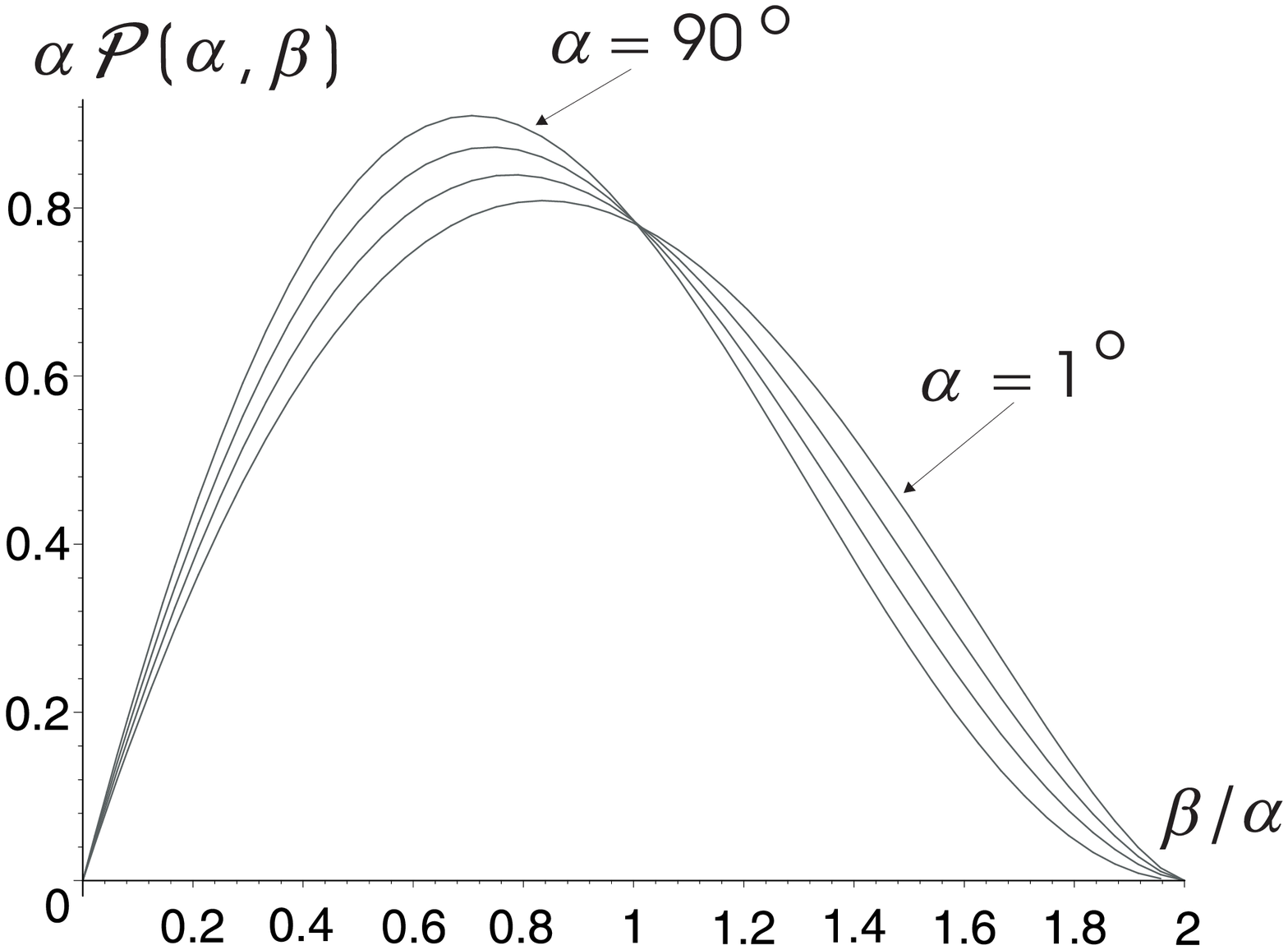,width=7cm,height=55mm}} 
\vspace*{3mm} 

\noindent {\small {\bf Bildo 2} {\it Denso de probablo $\P(\alpha,\beta)$, por pluraj fiksaj valoroj de} $\alpha$: $90^o, 75^o, 50^o, 1^o$.}  
\vspace*{5mm}

Ekvacio (\ref{Pbmea1}) estas skribata pli simple se ni difinas 
\bea                                                            \label{phis} %%14
\varphi:=\cos^{-1}\left(\frac{\cos\alpha}{\cos(\beta/2)}\right):                                                                   
\eea 
\vskip-2mm 
\bea     \nonumber                                                             
\P(\alpha,\beta)=2\kappa\sin\beta\Big[\sin^{-1}\left(\frac{(\cos\alpha+\cos\beta)\sin\varphi} {\sin\alpha(1+\cos\alpha)}\right) 
+(1-2\cos\alpha)\cos^{-1}\left(\frac{\sin(\beta/2)\cos\varphi}{\sin\alpha}\right)\Big].
\eea 
\vskip-12mm \bea \label{simples} { } \eea                                    %%15 

Nur kalotoj kun $\alpha<\pi/2$ evitas la ekvatoran regionon de lakta vojo. 
Sed tut\^ciela studoj jam okazis \cite{Evelise}, kun 
\bea                                                            \label{180} %%16
\P(\pi,\beta)=\frac{1}{2}\sin\beta. 
\eea

\section{Agnoskoj} 
Dankoj estas devataj al kolego Armando Bernui, por prezenti al ni la \^ci tiun problemon. 
Ni anka\w  \^suldas kolegon James R Piton, por amika helpo pri bibliografio en la reto, kaj al http://purl.org./NET/voko/revo/ por netaksebla lingva helpo.   
Informatikaj programoj Maple, Corel, PCTeX, kaj eps.fig, estis uzataj por fari \^ci tiun artikolon; ni volas kore danki al \^giaj kreintoj, same al ekipo de Microsoft kaj Arxiv/LANL.

\end{document}